\documentclass[fleqn,usenatbib]{mnras}
\usepackage{bm}
\usepackage[T1]{fontenc}
\DeclareRobustCommand{\VAN}[3]{#2}
\let\VANthebibliography\thebibliography
\def\thebibliography{\DeclareRobustCommand{\VAN}[3]{##3}\VANthebibliography}

\usepackage{graphicx}
\usepackage{amsmath}
\usepackage{amssymb}
\usepackage{xcolor}
\usepackage{ulem}

\usepackage{multirow}
\usepackage{booktabs}
\usepackage[flushleft]{threeparttable}
\usepackage{caption}
\def\lya{Lyman-$\alpha$}
\def\lyal{Lyman-$\alpha$ }

\def\hi{H$\,$\textsc{i} }
\def\his{H$\,$\textsc{i}}

\def\h2{${\rm H}_2$}
\def\h2s{${\rm H}_2$}
\def\tcmb{$T_{\rm CMB}$ }

\def\mathd{\mathrm{d}}

\def\nn{\nonumber}

\def\21cms{\textsc{21cmSense}}
\def\gmrt{\textsc{gmrt}}
\def\mwa{\textsc{mwa}}
\def\lofar{\textsc{lofar}}
\def\hera{\textsc{hera}}
\def\paper{\textsc{paper}}
\def\ska{\textsc{ska}}
\def\skalow{\textsc{ska1-low}}

\newcommand{\uni}[2]{{\rm #1}^{#2}} 

\definecolor{notecolor}{rgb}{0.8,0,0}

\title[21-cm power spectrum at $z<6$]{Implications of the $z>5$ Lyman-$\alpha$ forest for the 21-cm power spectrum from the epoch of reionization}

\author[Raste et al.]{{Janakee Raste$^{1}$\thanks{E-mail: janakee@theory.tifr.res.in},
  Girish Kulkarni$^{1}$,
  Laura C.~Keating$^2$,
  Martin G.~Haehnelt$^{3,4}$,}
\newauthor{Jonathan Chardin$^5$,
  Dominique Aubert$^5$}\\
  $^1$Tata Institue of Fundamental Research, Homi Bhabha Road, Mumbai 400005, India\\
  $^2$Leibniz Institute for Astrophysics Potsdam (AIP), An der Sternwarte 16, D-14482 Potsdam, Germany\\
  $^3$Institute of Astronomy, University of Cambridge, Madingley Road, Cambridge CB3 0HA, UK \\
  $^4$Kavli Institute of Cosmology, University of Cambridge, Madingley Road, Cambridge CB3 0HA, UK\\
  $^5$Observatoire Astronomique de Strasbourg, 11 rue de l'Universite, 67000 Strasbourg, France
}

\date{Accepted ---. Received ---; in original form ---}

\pubyear{2021}

\begin{document}
\label{firstpage}
\pagerange{\pageref{firstpage}--\pageref{lastpage}}
\maketitle

\begin{abstract}
  Our understanding of the intergalactic medium at redshifts
  $z=5$--$6$ has improved considerably in the last few years due to
  the discovery of quasars with $z>6$ that enable \lyal forest studies
  at these redshifts.  A realisation from this has been that hydrogen
  reionization could end much later than previously thought, so that
  large ``islands'' of cold, neutral hydrogen could exist in the IGM
  at redshifts $z=5$--$6$.  By using radiative transfer simulations of
  the IGM, we consider the implications of the presence of these
  neutral hydrogen islands for the 21-cm power spectrum signal and its
  potential detection by experiments such as \hera, \ska, \lofar, and 
  \mwa. In contrast with previous models of the 21-cm signal, we find
  that thanks to the late end of reionization the 21-cm power in our
  simulation continues to be as high as
  $\Delta^2_{21}=10~\mathrm{mK}^2$ at $k\sim 0.1~h/$cMpc at
  $z=5$--$6$.  This value of the power spectrum is several orders of
  magnitude higher than that in conventional models considered in
  the literature for these redshifts.  Such high values of the 21-cm
  power spectrum should be detectable by \hera\ and \skalow\ in $\sim 1000$
  hours, assuming optimistic foreground subtraction.  This
  redshift range is also attractive due to relatively low sky
  temperature and potentially greater abundance of multiwavelength
  data.
\end{abstract}


%%%%%%%%%%%%%%%%%%%%%%%%%%%%%%%%%%%%%%%%%%%%%%%%%%

%%%%%%%%%%%%%%%%% BODY OF PAPER %%%%%%%%%%%%%%%%%%

\section{Introduction}

Detecting the fluctuating 21-cm signal from neutral hydrogen during
the epoch of reionization is a major goal of extragalactic astronomy
in the coming decade.  Several experiments have published upper limits
on this signal \citep{GMRT2013, MWA2014, MWA2015, MWA2016, MWA2016b, LOFAR2017,
MWA2019, MWA2019b, PAPER2019, MWA2020, LOFAR2020, MWA2021} and some other
experiments are under development \citep{HERA2017, SKA2015a,
  SKA2015b}.
These upper limits have already been used to put constraints on the temperature and ionization state of the IGM \citep{2020MNRAS.493.4728G,2021MNRAS.503.4551G}. 

Most of these ongoing 21-cm experiments work under the assumption that
the 21-cm signal peaks when the universe is reionized at the 50\%
level and that the process of reionization itself ends at around
redshifts of $z\sim 6$, a number conventionally assumed due to early
constraints from the \lyal and Cosmic Microwave Background (CMB)
data. For example, the \hera\ EOR band is designed to operate at
frequencies 100--200~MHz which corresponds to redshifts 6.1 to
13.2. While it is acknowledged that the 21-cm signal has a large
uncertainty \citep{2017MNRAS.472.1915C}, this heuristic
understanding is used to set upper limits on the operating frequencies of the
radio experiments.

In recent years, however, large spatial fluctuations observed in
the Lyman-$\alpha$ forest at $z\approx 5.5$ have changed our
understanding of the end of the epoch of reionization \citep{Fan2006,
  2015MNRAS.447.3402B, 2018MNRAS.479.1055B, 2018ApJ...864...53E,
  2018ApJ...863...92B}.  The large scatter in the effective optical
depth $\tau_{\rm eff}$ of the \lyal forest at high redshift cannot be
explained due to only density or temperature fluctuations in a
post-reionization universe \mbox{\citep{2015ApJ...813L..38D}}.  It was also
found that the large \lyal absorption troughs observed by
\cite{2015MNRAS.447.3402B} and the rapid evolution of $\tau_{\rm eff}$
at redshifts 5--6 is difficult to explain using fluctuations in the
mean free path of ionizing photons due to spatial variation in the
photoionization rate \citep{2016MNRAS.460.1328D}.  It was proposed
that ultraviolet background fluctuations due to rare bright sources
such as quasars can explain the large fluctuations in \lyal forest
$\tau_{\rm eff}$ \citep{2015MNRAS.453.2943C, 2017MNRAS.465.3429C}, but
the abundance of QSOs at high redshifts is probably too low
\citep{2019MNRAS.488.1035K}.

Using radiative transfer simulations, \cite{2019MNRAS.485L..24K}
suggested that the observed large spatial scatter in the
Lyman-$\alpha$ forest opacity can arise due to large regions, with 
sizes of about 100 comoving Mpc, of neutral hydrogen during the last
stages of reionization \citep{2020MNRAS.491.1736K}.  Due to these
``neutral islands'', which persist in the in low-density regions of the
Universe upto $z\approx 5.5$, the end of the reionization is delayed
to $z\sim 5.3$ \citep{2020MNRAS.494.3080N,2021MNRAS.501.5782C,2021arXiv210109033Q}.
This
late reionization model is able to reproduce the distribution of
$\tau_{\rm eff}$ of the Lyman-$\beta$ forests at $4<z<6$
\citep{2020MNRAS.497..906K}.  This model is also consistent with other
data, such as the underdensity of Lyman-$\alpha$ emitters (LAEs; \citealt{2018ApJ...863...92B}) and Lyman-Break Galaxies (LBGs; \citealt{2020ApJ...888....6K})
in the vicinity of long \lyal absorption troughs
\citep{2020MNRAS.491.1736K}, the luminosity and angular correlation
functions of LAEs \citep{2019MNRAS.485.1350W}), measurements of the Thomson scattering
optical depth to the last scattering surface \citep{2019MNRAS.485L..24K},
and constraints on the
ionized hydrogen fraction from quasar damping wings \citep{2015MNRAS.447..499M, 2017MNRAS.466.4239G, 2018ApJ...856....2M, 2018ApJ...864..142D, 2019MNRAS.484.5094G, 2020ApJ...896...23W}.

A delayed reionization history would be consequential for 21-cm
  experiments as it would shift the target 21-cm signal to higher
  frequencies.  In this paper, we discuss the implications of a late
  end to the epoch of reionization and a persistence of neutral
  hydrogen islands at $z<6$ for the 21-cm power spectrum.  We use the
  radiative transfer simulation presented by
  \citet{2019MNRAS.485L..24K}, examine the power spectrum of the 21-cm
  signal at $z<6$ and the prospects of detecting it.

We assume a flat $\Lambda$CDM universe with
$\Omega_\mathrm{b}=0.0482$, $\Omega_\mathrm{m}=0.308$,
$\Omega_\Lambda=0.692$, $h=0.678$, $n_\mathrm{s}=0.961$,
$\sigma_8=0.829$, and $Y_\mathrm{He}=0.24$
\citep{2014A&A...571A..16P}.  The units `ckpc' and `cMpc' refer to
comoving kpc and comoving Mpc, respectively.  In
Section~\ref{sec:methods}, we give details of our simulation and our
model of the 21-cm signal. Section~\ref{sec:results} presents the
21-cm power spectrum in our model at various redshifts. In
Section~\ref{sec:detection}, we discuss the prospects of detecting
this signal with four current and upcoming interferometric experiments
(\mwa, \lofar, \hera\ and \skalow).  We end with a discussion in
Section~\ref{sec:conclusion}.

\section{Methodology} \label{sec:methods}

\subsection{21-cm Signal} \label{sec:21cm}

In the high-redshift universe, neutral hydrogen (\his) is the most
abundant element. The hyperfine transition of the \hi ground state
corresponds to a photon of frequency $\nu_{21} = 1420.406$~MHz (which
corresponds to wavelength $\lambda_{21} = 21.1 \, \rm cm$). The
distortion in the CMB spectrum due to this transition contains
information about the density, temperature and ionization of the
\hi gas \citep{1997ApJ...475..429M, 1999A&A...345..380S,
  2004ApJ...608..611G, Sethi05}. This distortion is a function of the
ratio of the number of atoms in the two hyperfine states, $n_1/n_0 =
(g_1/g_0)\; \mathrm{exp}(-h_p \nu_{21}/k_B T_{\rm S})$, where $n_0$ ($n_1$)
and $g_0=1$ ($g_1 = 3$) are the number density and the degeneracy of
atoms in the singlet (triplet) hyperfine state, and $h_p$ is the Planck
constant. Note that the spin temperature $T_\mathrm{S}$ is the thermal
temperature the \hi gas would have if the number densities of singlet
and triplet states were in thermal equilibrium.  The spin temperature
is determined by the detailed balance between various processes that
can alter the level population of the \hi hyperfine states
\citep{Field1958,21cm_21cen}, and is given by
\begin{align}
  T_{\rm S}^{-1}=\frac{T_{\gamma}^{-1}+x_{\alpha}T_{\alpha}^{-1}+x_c T_{\rm K}^{-1}}{1+x_{\alpha}+x_c}.
  \label{eq:TS_x}
\end{align}
The collision coefficient, $x_c$, is proportional to the number densities
of electrons $n_e$ and neutral hydrogen atoms $n_{\rm HI}$, and
increases with the kinetic temperature $T_{\rm K}$. The \lyal coupling
coefficient $x_\alpha$ is proportional to the number density of \lyal
photons. Such photons are produced in sources within the
galaxies as well as by the interaction of X-ray photons with the IGM
\citep{Heating2001}. Repeated scattering of these \lyal photons by \hi
gas at kinetic temperature $T_{\rm K}$ causes the photon colour
temperature $T_\alpha$ to relax to $T_{\rm K}$ through the Wouthuysen-Field
effect \citep{Wouthuysen1952,Field1958}.  Therefore, if
$x_{\text{tot}}=x_c+x_{\alpha} \gtrsim 1 $, then $T_{\rm S}$ is strongly
coupled to $T_{\rm K}$.

When the background CMB radiation passes through a cloud
of \hi with spin temperature $T_{\rm S}$, 21-cm photons are
absorbed from the blackbody spectrum if $T_{\rm S} < T_{\rm CMB}$ or emitted
into it if $T_{\rm S} > T_{\rm CMB}$. The observed change in CMB brightness
temperature caused by a cloud at redshift $z$ is,
\begin{align}
  \Delta T_b(\nu_o) = \frac{T_{\rm S}(z) - T_{\rm CMB} (z)}{1+z} (1-\mathrm{e}^{-\tau}), \label{eq:21Tb1}
\end{align} 
where $\nu_o = \nu_{21}/(1+z)$ is the observed frequency, $\tau = \int
\alpha_\nu \mathrm{d}s$ is the 21-cm optical depth of the cloud, and
$\alpha_\nu$ is the absorption coefficient. Since the natural
broadening of the 21-cm line is very small,
%(because $A_{21} = 2.85 \pow{-15} \uni{s}{-1}$ is very small),
its resonance line width is the Doppler width dominated by the motion of the atoms,
\begin{align}
    \mathd \nu = \nu \frac{\mathd v}{c} = \nu \frac{\mathd s}{c} \left[ H(z) + \frac{\mathd v_p}{\mathd s} \right].
\end{align}
Here, $v_p$ is the component of the peculiar velocity of the gas
parallel to our line of sight and $s$ is the light-travel distance.
In our simulation, the comoving size of each cell is $\mathd r =
160/2048~{\rm cMpc}/h$ (see section~\ref{sec:simulation}) and the
light-travel distance between two cells is $\mathd s \simeq \mathd
r/(1+z)$.  We can safely assume that there is no significant evolution
of cosmological quantities over this distance.  Therefore, assuming a top-hat \hi
line profile function ($\phi(\nu) \simeq 1/\Delta \nu$), the
optical depth is
\begin{equation}
\tau = \alpha_\nu \Delta s \propto \frac{c}{\nu} \left[H(z) + \frac{\mathd v_p}{\mathd s} \right]^{-1} .
\end{equation}
During the matter-dominated epoch, the optical depth is then
\begin{align}
  \tau
  &\simeq \frac{3^2}{2^8 \pi^2 } \frac{A_{21}h_p c^3}{G m_p k_B \nu^2} \frac{H_0}{h} \frac{x_{\rm HI} (1+\delta)(1+z)^{3/2}}{T_{\rm S} \left(1 + 1/H(z) (\mathd v_p/\mathd s)\right) } \frac{Y_{\rm H} \Omega_b h^2}{(\Omega_m h^2)^{1/2}} \nn \\
  &\simeq x_{\rm HI} (1+\delta) \left(\frac{8.55\; {\rm mK}}{T_{\rm S}}\right) \left(1+\frac{1}{H(z)} \frac{\mathd v_p}{\mathd s}\right)^{-1}(1+z)^{3/2}\nn\\
  & \qquad \times \left(\frac{Y_H}{0.76}\right)  \left(\frac{\Omega_b h^2}{0.022}\right) \left(\frac{0.14}{\Omega_m h^2}\right)^{1/2}.
  \label{eq:21tau3} 
\end{align}
When the peculiar velocity gradient is negligible compared to the Hubble flow ($|\mathd v_p/\mathd s| \ll H(z)$), this optical depth is small ($\tau \ll 1$), and we can approximate $(1-\mathrm{e}^{-\tau}) \simeq \tau$ in Equation~(\ref{eq:21Tb1}). 
In our simulation, most grid cells satisfy the condition $|\mathd v_p/\mathd s| \ll H(z)$. For the small number of cells with larger peculiar velocity gradient, we apply a cutoff on the peculiar velocity gradient $|\mathd v_p/\mathd s| \leq 0.5 H(z)$ following previous literature \citep{2010MNRAS.406.2421S,21CMFAST,2012MNRAS.422..926M}. 
As our focus in this paper is on the end of reionization, at lower redshifts, we expect the effect of redshift space distortions to be small \citep{2013MNRAS.435..460J,2014MNRAS.443.2843M}.

Finally, the differential brightness temperature is,
\begin{align}
  \Delta T_b(\nu_o) &\simeq \tau\; \frac{T_{\rm S}(z) - T_{\rm CMB} (z)}{1+z}  \nonumber \\
  &\simeq 27 \; {\rm mK}\; x_{\rm HI} (1+\delta) \left(1-\frac{T_{\rm CMB} (z)}{T_{\rm S}}\right)\nn\\
  &\qquad\times\left(1 + \frac{1}{H(z)} \frac{\mathd v_p}{\mathd s}\right)^{-1}\left(\frac{1+z}{10}\right)^{1/2}\nn\\
  &\qquad\times\left(\frac{Y_H}{0.76}\right)\left(\frac{0.14}{\Omega_m h^2}\right)^{1/2} \left(\frac{\Omega_b h^2}{0.022}\right).
  \label{eq:21Tb}
\end{align} 
Here, $\Delta T_b$ is a function of both the redshift of observation (through $\nu_o$) and the direction of observation. Fluctuations in \hi density ($\delta$), ionization state ($x_{\rm HI}$), spin temperature ($T_{\rm S}$) and peculiar velocity gradient ($\mathd v_p/\mathd s$) contribute to the spatial fluctuations of $\Delta T_b$ at any redshift.

In our work, we assume that at $z \leq 10$, (\textit{a}) the spin temperature is strongly coupled to the kinetic temperature through \lyal coupling ($T_{\rm S}=T_{\rm K}$), and (\textit{b}\/) the gas is sufficiently heated ($T_{\rm K} \gg T_{\rm CMB}$). Then Eq.~\ref{eq:21Tb} can be simplified as, 
\begin{align}
  &\Delta T_b(\nu_o) \simeq 27 \; {\rm mK}\; x_{\rm HI} (1+\delta)  \left(1 + \frac{1}{H(z)} \frac{\mathd v_p}{\mathd s}\right)^{-1}\nn\\
  & \qquad\times\left(\frac{1+z}{10}\right)^{1/2}\left(\frac{Y_H}{0.76}\right)\left(\frac{0.14}{\Omega_m h^2}\right)^{1/2} \left(\frac{\Omega_b h^2}{0.022}\right).
\label{eq:21Tb_2}
\end{align}
In this case, the global 21-cm signal is always in emission ($ \langle \Delta T_b \rangle > 0$) and small ($\langle \Delta T_b \rangle < 28$~mK at $z < 10$).
The only sources of fluctuations are density, ionization and line-of-sight peculiar velocity gradient. The assumption $T_{\rm K} \gg T_{\rm CMB}$ is a plausible approximation at redshifts of our interest, since heating due to X-rays from first sources is expected to raise temperature of the neutral gas well above the $T_{\rm CMB}$ by $z \sim 10$ \citep{EDGES2018,2021MNRAS.503.4551G}. 
Some recent models have suggested that X-ray heating might be delayed to $z<10$ \citep{LateHeat2,2017ApJ...840...39M,2017MNRAS.464.1365M,2017MNRAS.472.1915C,2020MNRAS.491.3891P,2020MNRAS.493.1217M}. Our assumption ($T_{\rm S} \gg T_{\rm CMB}$) will not be valid in such scenarios, but other than the redshift of heating transition ($T_{\rm K} \simeq T_{\rm CMB}$), the spin temperature fluctuations are only expected to enhance the power spectrum of 21-cm signal \citep{LateHeat2, 2014MNRAS.439.3262M, RS18, RS19,2021ApJ...912..143M}.

\begin{figure*}
  \includegraphics[width=\linewidth]{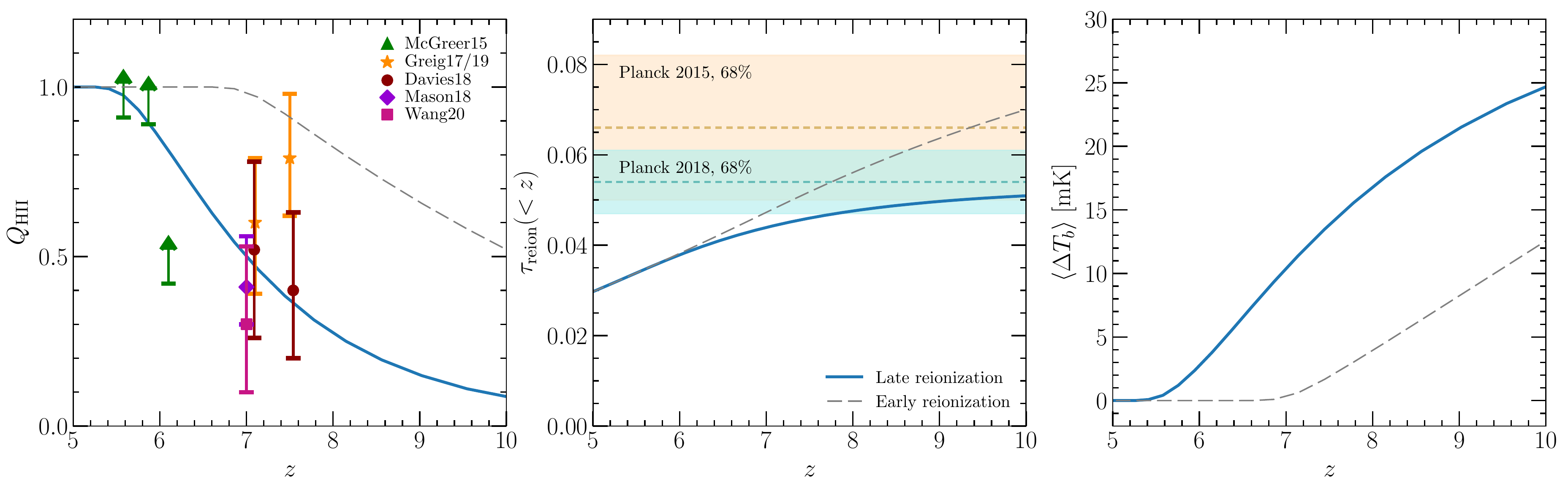}
  \caption{Evolution of the ionized-hydrogen volume fraction $Q_\mathrm{HII}$ (left panel), cumulative CMB Thomson scattering optical depth $\tau_{\rm reion}$ (middle panel) and the global average of the 21-cm differential brightness temperature $\langle\Delta T_b\rangle$ (right panel) in our fiducial simulation (blue curves). Reionization is completed by $z \sim 5.3$ in our model. For comparison we show an early reionization model as the grey dashed curves. We have overlaid constraints on the ionization fraction from \lyal absorption studies of quasar spectra \protect\citep{2015MNRAS.447..499M, 2017MNRAS.466.4239G, 2018ApJ...856....2M, 2018ApJ...864..142D, 2019MNRAS.484.5094G, 2020ApJ...896...23W}.  The middle panel shows $\tau_{\rm reion}$ measurements from \protect\cite{Planck2015} and \protect\cite{Planck2018}.  The right panel assumes $T_{\rm S} \gg T_{\rm CMB}$.}
  \label{fig:ave_all}
\end{figure*}

\begin{figure*}
  \includegraphics[width=\linewidth]{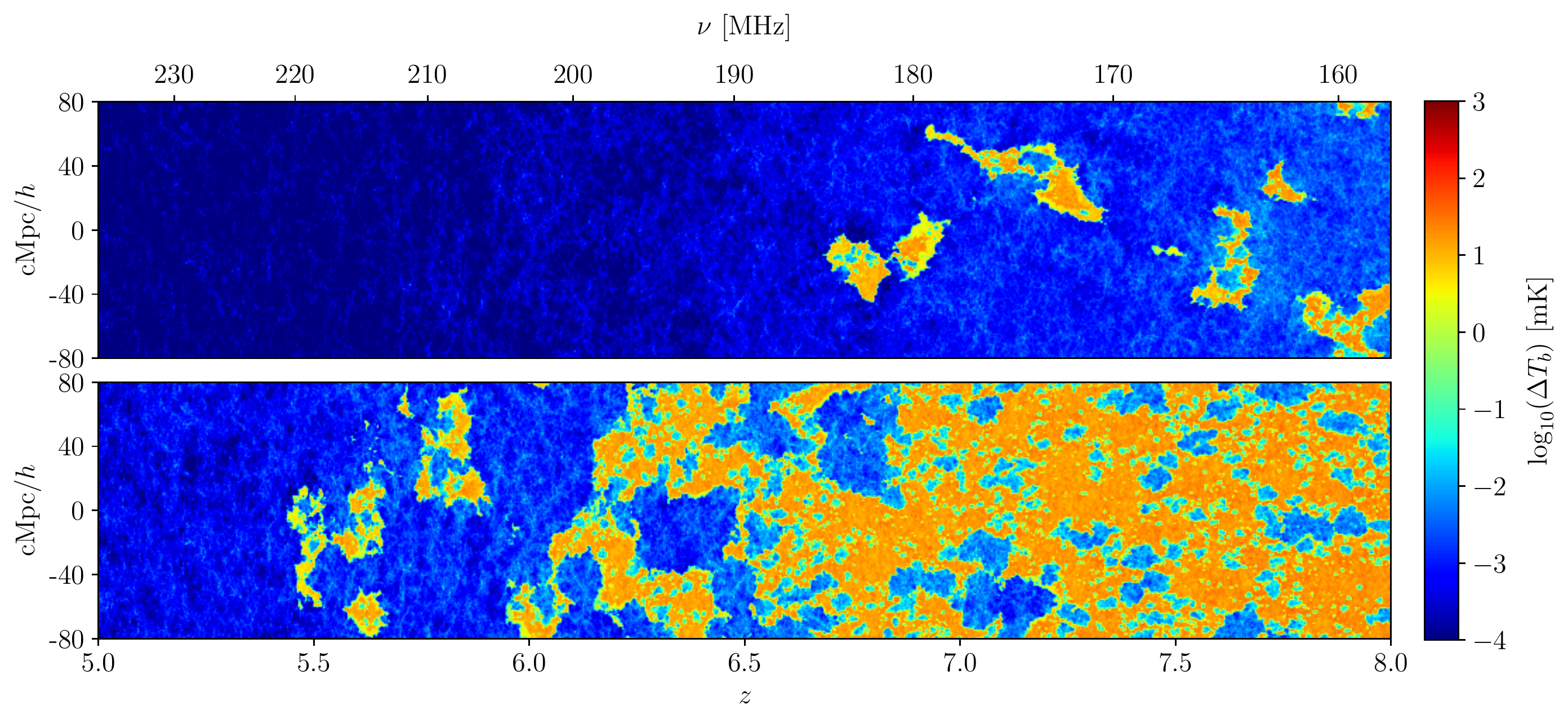}
  \caption{Brightness temperature ($\Delta T_b$) lightcones of early (top panel) and late (bottom panel) reionization models from redshift 8 to 5. In the early model, reionization is completed by $z \sim 6.7$ and the IGM is highly ionized by redshift 5.5. In the late reionization model that is preferred by the $z>5$ Ly~$\alpha$ data, there are large neutral regions present in the IGM at this redshift, which produce a large 21-cm power spectrum signal.}
  \label{fig:lc}
\end{figure*}

\subsection{Simulation} \label{sec:simulation}

Our fiducial reionization model is that presented by
\cite{2019MNRAS.485L..24K}.  We give the essential details here for
completeness.  The simulation is performed in two stages.  In the
first stage, a cosmological hydrodynamical simulation is performed in
order to obtain the gas density and velocity fields.  In the second
stage, the radiative transfer is computed to obtain the distribution
of the neutral hydrogen during the epoch of reionization.  We used the
\textsc{p-gadget-3} code, a modified version of the \textsc{gadget-2}
code \citep{2001NewA....6...79S, 2005MNRAS.364.1105S},  for the
hydrodynamic simulation.  Our box size is $160\,h^{-1}\,\text{cMpc}$,
with periodic boundary conditions, and there are $2048^3$ gas and dark
matter particles. The simulation is evolved from $z=99$ down to $z=4$.
Snapshots of gas density, halo masses, and other quantities are saved
at 40~Myr intervals.  This hydrodynamical simulation is similar to the
simulations from the Sherwood Simulation Suite
\citep{2017MNRAS.464..897B}.  Our initial conditions are identical to
the initial conditions used in their 160--2048 simulation.

The mean free path of ionizing photons is set by self-shielded regions
with a typical overdensity of $\Delta=10$--$100$
\citep{2009MNRAS.394.1812P, 2018MNRAS.478.1065C}.  Therefore, it is
safe to use the {\tt QUICK\_LYALPHA} option in
\mbox{\textsc{p-gadget-3}}, which simplifies galaxy formation and
speeds up the simulation by converting gas particles with temperature
less than $10^5$\,K and overdensities greater than $1000$ to star
particles \citep{2004MNRAS.354..684V} and removing them from the
hydrodynamical calculation.

In order to improve the accuracy of the small-scale hydrodynamics,
heat is injected in the simulation box to obtain a realistic pressure
smoothing at lower redshifts.  To accomplish this, instantaneous
reionization is assumed at redshift $z=15$ and ionization equilibrium
with the metagalactic UV background is modelled according to a
modified version of \cite{2012ApJ...746..125H}
reionization model. This yields IGM temperatures that agree with
measurements by \cite{2011MNRAS.410.1096B}.  The pressure smoothing
scale at redshifts $z>5$ for this UV background is less than
$100h^{-1}\,\text{ckpc}$ \citep{2015ApJ...812...30K,
  2017ApJ...837..106O}, which is approximately equal to the cell size
of our grid ($78.125h^{-1}\,\text{ckpc}$, described below).
Therefore, the absence of the coupling between the radiative transfer
and the gas hydrodynamics does not significantly affect our
computation of \lya\ opacities from the simulation for calibration.

The radiative transfer is computed using the \textsc{aton} code
\citep{2008MNRAS.387..295A, 2010ApJ...724..244A}.  We grid the gas
density by projecting the smooth particle hydrodynamic (SPH) kernel in
our simulation onto a Cartesian grid. The number of grid cells is set
to be equal to the number of gas particles in the simulation
($2048^3$).  This gives a grid resolution of
$78.125\,h^{-1}\,\text{ckpc}$.  Haloes that host our ionization
sources are identified using the friends-of-friends algorithm.  At
$z=7$, this yields a minimum halo mass of $2.3\times
10^{8}h^{-1}\,\text{M}_\odot$, which is close to the atomic hydrogen
cooling limit, and the maximum halo mass is $3.1\times
10^{12}h^{-1}\,\text{M}_\odot$.  We place sources of ionizing
radiation at the centres of in haloes with masses greater than
$10^9\,$M$_\odot$ as the halo mass function below this mass suffers
from incompleteness due to limited resolution of the simulation.
\textsc{aton} solves the radiative transfer equation by using the M1 approximation \citep{2008MNRAS.387..295A, 1984JQSRT..31..149L, 2008ASPC..385...91G}
for the first moment.  A halo with mass $M$ is assumed to emit
hydrogen-ionizing photons with a rate $\dot N=\alpha M$ and the
average ionizing photon emissivity of the total simulated volume is
$\dot n = \alpha \sum M / V_\mathrm{box}$, where $V_\mathrm{box}=L^3=
160^3~({\rm cMpc}/h)^3$ is the simulation box volume and the sum is
over all haloes which host sources.  The parameter $\alpha$ is the
only quantity that is varied in order to calibrate the simulation to
given observations, such as the \lya\ forest
\citep{2019MNRAS.485L..24K}.  It is assumed to be a function of
redshift, but independent of halo mass and it encodes the details of
the astrophysical processes such as star formation and photon escape
through the inter-stellar medium which govern the ionizing photon
production in galaxies.  
Our assumed scaling between the ionizing luminosity and the halo mass
is related to a scaling relation between the observed UV luminosity of
high-redshift sources and the halo mass via the unknown ionizing
escape fraction.  For a mass-independent escape fraction, as a result,
we have $L_\mathrm{UV}\propto M$.  This yields a reasonably good fit
to the observed high-redshift galaxy luminosity functions \citep{2015MNRAS.453.2943C}, although an even better fit may be obtained
if the scaling between $L_\mathrm{UV}$ and $M$ is made slightly
steeper at low halo masses ($M\lesssim 10^{11}$~M$_\odot$) and flatter
at higher halo masses \citep{2010ApJ...714L.202T}.  But, as the escape
fraction is expected to decrease with halo mass \citep{2017MNRAS.466.4826K}, this modified scaling can easily be absorbed in the
halo-mass dependence of the escape fraction to yield consistency between
our assumed ionizing emissivity and the observed UV luminosity
function.
All sources are assigned a blackbody spectrum
with $T=70\,000\,\text{K}$ \citep{2018MNRAS.477.5501K}, which
corresponds to an average photon energy of 23.83\,eV in the
optically-thick limit.  A single photon frequency is used for the
radiative transfer in order to reduce computational cost.  The
reionization history is not significantly affected by choice of these
parameters, because the simulation is calibrated to match with the the
\lya\ forest data. Therefore, any change in the gas temperature due to
changing the source spectrum can then be compensated by changing
$\alpha$ in the source emissivity above.  Finally, we use
Equation~\ref{eq:21Tb_2} to calculate differential brightness
temperature ($\Delta T_b$) box using the density, ionization, and
peculiar velocity boxes. We arbitrarily take the $z$-axis of the
simulation box as the line-of-sight direction and calculate the
peculiar velocity gradient along this axis.  If the peculiar velocity
gradient for any cell $\mathd v_p/\mathd s < -0.5 H(z)$ ($\mathd
v_p/\mathd s > 0.5 H(z)$), then we set its value to $-0.5 H(z)$ ($0.5
H(z)$) \citep{2010MNRAS.406.2421S,21CMFAST}.

In this work we also consider a second radiative transfer simulation
in which the evolution of the volume-averaged ionized hydrogen
fraction is calibrated to match its evolution in the
\cite{2012ApJ...746..125H} model of reionization.  In this model,
reionization is complete at $z\sim 6.7$.  The calibration is achieved
by adjusting the source emissivity in the simulation at each time step
to get the desired ionized fraction evolution.  The two simulations
are identical in all aspects apart from the source emissivity.  In
this paper, we refer to the \cite{2012ApJ...746..125H} model of
reionization as `early reionization model', while referring to our fiducial
simulation \citep{2019MNRAS.485L..24K} as the `late reionization
model'.

\section{Results} \label{sec:results}
\begin{figure*}
  \includegraphics[width=0.99\textwidth]{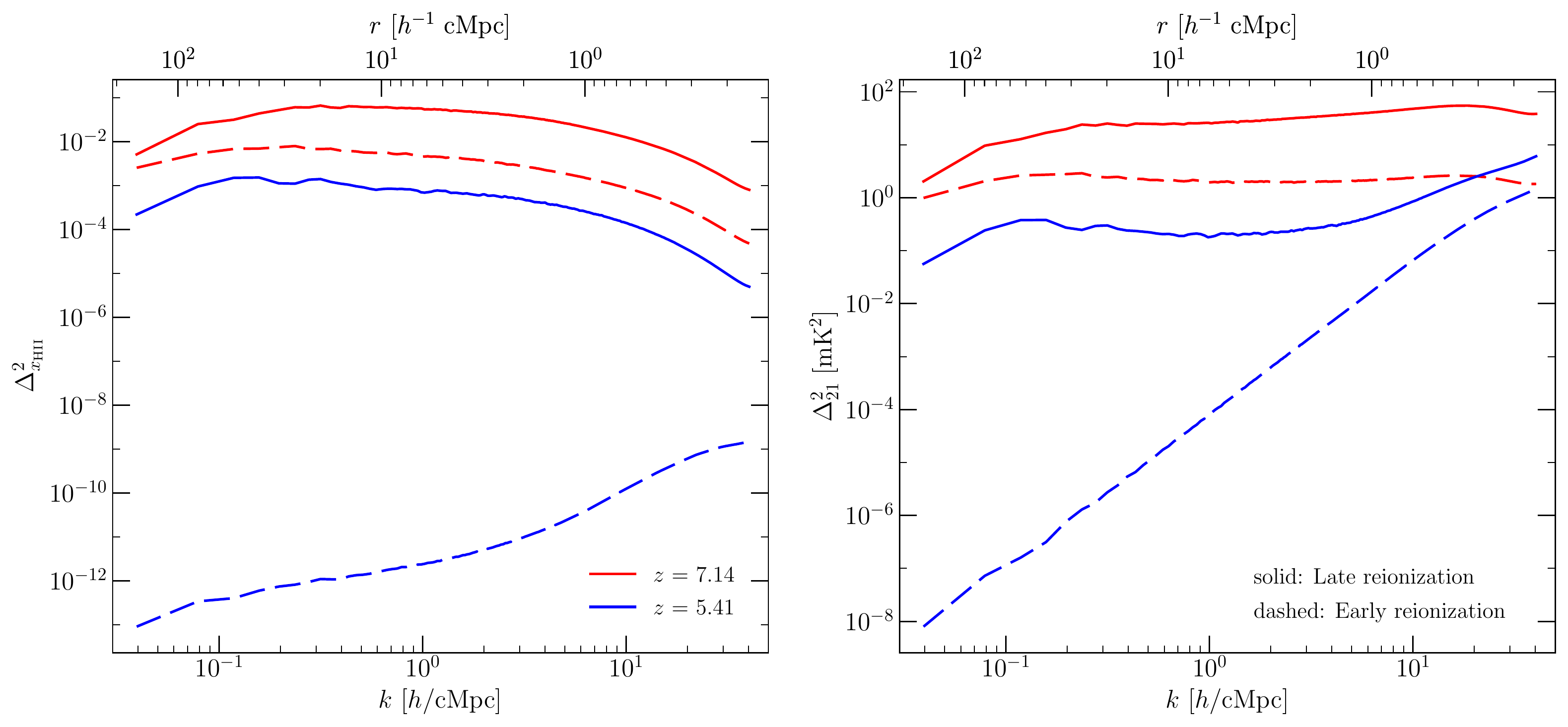}
  \caption{Power spectrum of the ionized-hydrogen fraction (left
    panel) and the 21-cm brightness temperature ($\Delta T_b$) (right
    panel) at $z=5.41$ and $z=7.14$ in our early and late reionization
    models.  At $z<6$ the late reionization model, preferred by the
    Ly~$\alpha$ data, shows orders of magnitude greater power than the
    more conventional early reionization model.  In the late
    reionization model, the ionization power spectrum peaks at
    $z=7.14$, which approximately corresponds to the mid-point of
    reionization (Figure~\ref{fig:ave_all}). The power spectrum
    decreases at lower redshifts in this model. We also note that with
    time the peak of the power spectra shifts to smaller $k$ (larger
    $r$) due to the growth of the ionized regions. The brightness
    temperature fluctuations are due to both the density and
    ionization fluctuations. On large scale (small $k$), the $\Delta
    T_b$ fluctuations show a peak at roughly the same scale where
    ionization fluctuations have a peak. However, at small scales
    (large $k$), they have a shape similar to the density
    fluctuations.}
  \label{fig:PS_all2}
\end{figure*}

% ------------------- AVERAGE -------------------

We show the ionization history of our model in Figure~\ref{fig:ave_all}. As discussed by \cite{2019MNRAS.485L..24K}, the presence of self-shielded neutral hydrogen islands at low redshift delays the end of reionization, which is completed by $z \sim 5.3$. We compare our results with the early reionization model of \cite{2012ApJ...746..125H}. In this model, the volume averaged ionization fraction reaches 0.5 at $z\sim 10$ and the reionization ends at $z\sim 6.7$. 
In our late reionization model, the mid-point of reionization is delayed to $z\sim 7$; this is in agreement with the inference from \lyal absorption studies of quasar spectra at $5<z<8$ \citep{2015MNRAS.447..499M, 2017MNRAS.466.4239G, 2018ApJ...856....2M, 2018ApJ...864..142D, 2019MNRAS.484.5094G, 2020ApJ...896...23W}. 

In the middle panel of Figure~\ref{fig:ave_all}, we show the Thomson scattering optical depth, $\tau_{\rm reion} = 0.054 \pm 0.007$, by \citet{Planck2018} as a blue shaded region. Previous measurements of $\tau_{\rm reion}$ predicted an earlier epoch of reionization. For example, the orange shaded region in the figure corresponds to the $\tau_{\rm reion} = 0.066 \pm 0.016$, given by \cite{Planck2015}. 
The integrated Thomson scattering optical depth of the reionization history of our late reionization model is compatible with the latest measurement of  $\tau_{\rm reion}$. The early reionization model predicts a higher $\tau_{\rm reion}$, which disagrees with \cite{Planck2018}, but was consistent with earlier measurements of $\tau_{\rm reion}$. 

Figure~\ref{fig:ave_all} also shows the evolution of the globally averaged 21-cm signal in our simulations.  As discussed in the previous section, we assume that the \hi spin temperature $T_{\rm S} \gg$ \tcmb at $z \leq 10$.  As a result, our globally averaged 21-cm differential brightness temperature is small and positive ($0\leq \langle \Delta T_b \rangle \leq 28$) (Equation~\ref{eq:21Tb_2}). With the progress of reionization, the brightness temperature decreases.
%
% ------------------ LIGHTCONE --------------------
%
We show the brightness temperature lightcones for these two models in Figure~\ref{fig:lc} at redshifts 5--8. The blue regions, which correspond to low brightness temperature, are ionized, whereas the orange regions corresponding to large brightness temperature are neutral. The ionized regions have already overlapped by redshift 8 in the early reionization model, and the regions of substantial brightness temperature disappear by redshift 6.7. At lower redshifts, the brightness temperature structure is well below detection level (see discussion in following section). In the late reionization model, the Universe is mostly neutral at $z=8$. The ionized regions grow at lower redshifts and large regions ($\sim 100$~cMpc) of substantial brightness temperature are still present in the IGM at redshifts as low as 5.5.

% -------------- 1D POWER SPECTRUM -----------------
\begin{figure}
  \includegraphics[width=\columnwidth]{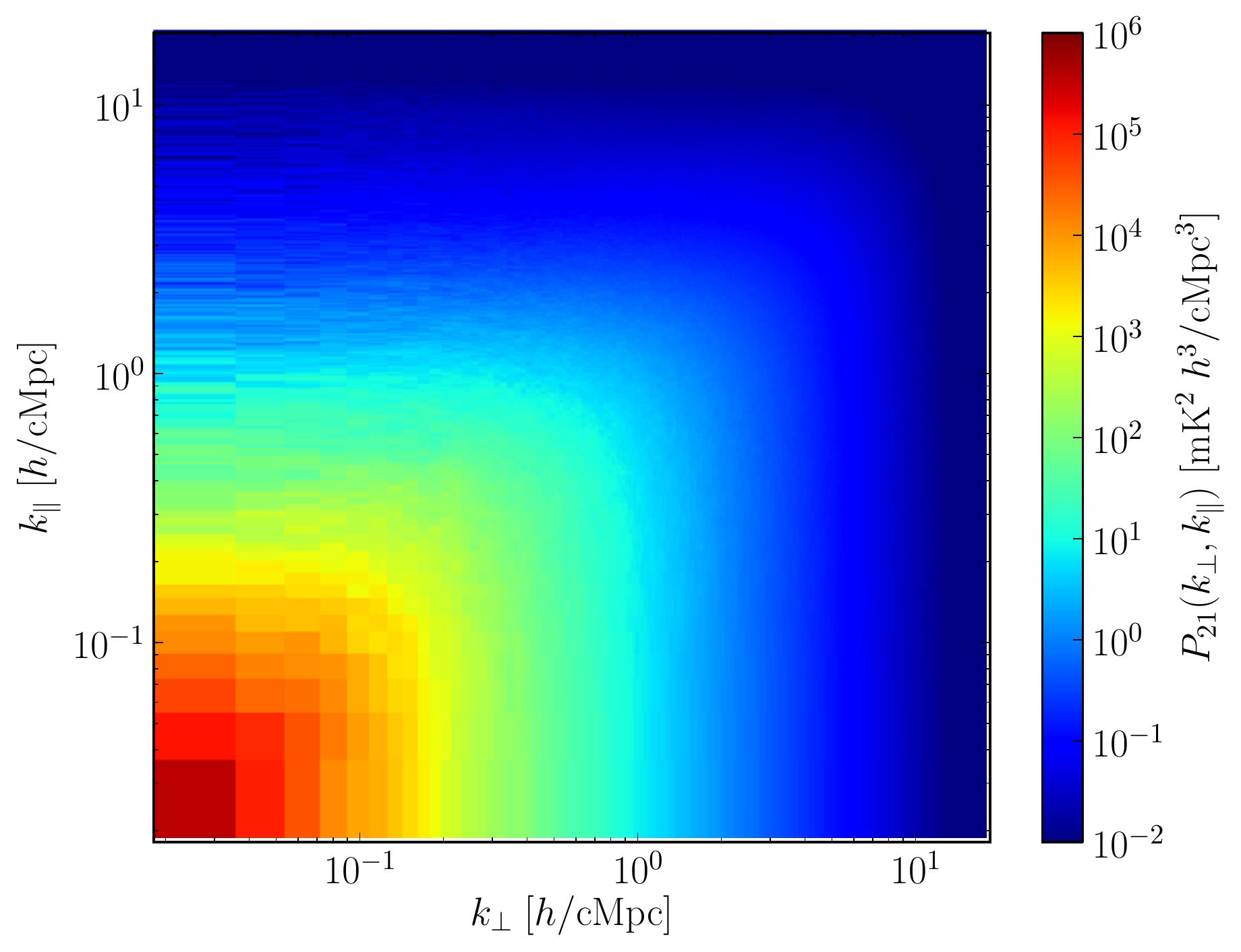}
    \caption{Two-dimensional cylindrically-averaged 21-cm power spectrum at $z=5.58$ in our model.}
  \label{fig:PS_kk_22}
\end{figure}

We calculate the 1D spherically averaged power spectrum of a statistically isotropic quantity $F(\bm{r})$ as, 
\begin{equation}
  \langle \Tilde{F}(\bm{k_1}) \Tilde{F}(\bm{k_2}) \rangle = (2 \pi)^2 \delta_D (\bm{k_1}+\bm{k_2})P_F(k),
  \label{eq:PS}
\end{equation}
where, $\Tilde{F}(\bm{k})$ is the Fourier transform of $F(\bm{r})$, $\delta_D$ is the Dirac delta function and angular brackets denote ensemble averages. Assuming ergodicity, we take a volume average over $k$-modes in all directions. 
Figure~\ref{fig:PS_all2} shows 
\begin{align}
    \Delta_F^2(k) = \frac{k^3}{2\pi^2} \frac{\langle \tilde{F}^2(k)\rangle}{V_{\rm box}}
\end{align}
of the ionized hydrogen fraction and the 21-cm brightness temperature at $z=5.41$ and $z=7.14$ in our early and late reionization models. Here, $V_{\rm box} = 160^3~({\rm cMpc}/h)^3$ is the volume of the simulation box.
In our fiducial late reionization model we see that both the ionization and brightness temperature power spectra drop as we go from $z = 7.14$ to $z = 5.41$. This is because redshift 7.14 is close to the midpoint of reionization ($ Q_{\rm HII} \simeq 0.5$) in our model (Figure~\ref{fig:ave_all}). A similar drop is also seen in the early reionization model. However, due to the early end of the reionization in this model, both the ionization and brightness power spectra at redshift 5.41 are orders of magnitude smaller than their counterparts in the late reionization model.

Further, we can also notice in Figure~\ref{fig:PS_all2} that the ionization power spectrum peaks at different scales in the two models at redshifts 7.14 and 5.41.
The peak of the ionization power spectrum shifts to smaller $k$ (larger $r$) with time, corresponding to the growth of ionized regions. In the late reionization model, the peak of the power spectrum at $z=7.14$ is at $k\sim 0.9~h/{\rm cMpc}$ and shifts to $k\sim 0.6~h/{\rm cMpc}$ at $z=5.41$. 
%As the ionized regions in the early reionization model grow large earlier compared to our late reionization model, we observe that the peak of ionization power spectrum is at smaller $k$ (larger $r$) for the early reionization model at $z = 7.14$. 
At small $k$ (large $r$), the $\Delta T_b$ power spectra have a shape similar to the ionization power spectra, while at large $k$ (small $r$), they have a shape similar to the matter density power spectra. 
The shape and amplitude of our power spectra at the midpoint and end of reionization broadly match with the literature \citep{2011MNRAS.414..727Z,2014ApJ...789...31L,Ghara2015,2018MNRAS.477.1549H,2021MNRAS.503.3698H,2021MNRAS.504.2443B}.

% -------------- 2D POWER SPECTRUM -----------------

Redshift-space distortions caused by the line-of-sight peculiar velocities will introduce anisotropy in the $\Delta T_b$ power spectra between $k$-modes which are parallel ($k_\parallel$) and perpendicular ($k_\perp$) to the line-of-sight. 
We compute the 2D cylindrical power spectra $P_{21}(k_\perp, k_\parallel) = \langle \Tilde{T}_b^2(k_\perp, k_\parallel) \rangle/V_{\rm box}$, where we separately average over the $k_\perp$ (along z-axis) and $k_\parallel$ (in xy-plane) modes. We show $P_{21}(k_\perp, k_\parallel)$ at $z=5.58$ in Figure~\ref{fig:PS_kk_22} for our late reionization model. As seen in Figure~\ref{fig:PS_all2}, the $\Delta T_b$ fluctuations are dominated by ionization fluctuations when the ionization fraction is large, and the effect of peculiar velocity is small at these redshifts \citep{2013MNRAS.435..460J,2014MNRAS.443.2843M}.

% NOTE: k_perp on X-axis and k_par on Y-axis

\section{Prospects of Detection}\label{sec:detection}

\begin{table*}
%	\centering
	\begin{threeparttable}
	\caption{Summary of observational parameters of four instruments: \mwa, \lofar, \hera\ and \skalow. The number of antenna elements are given as core antennas + remote/outrigger antennas. The effective collecting area $A_e$ is measured at 150~MHz at the zenith. The full-width at half maximum (FWHM), the field of view, and the angular resolution are also given for 150~MHz.}
	\label{tab:Params}
    \begin{tabular}{ccccccc} 
	\toprule
		Parameter									&\mwa$^{\rm a}$ [High] [Low]& \lofar$^{\rm b}$ HBA 	& \hera$^{\rm c}$ [EoR]	& \skalow$^{\rm d}$		\\
		\hline
		\hline
        Number of antennae, $N_a$ 					& 128 + 128		& 48 + 22 		& 320 + 30		& 224 + 288 	\\
        Number of antennae used in \21cms			& 128 			& 48	 		& 331 			& 224 	        \\

        \hline
        Core radius $r_{\rm core}$ [m] 				& 300		 	& 2000			& 150 			& 500 			\\
        Maximum radius, $r_{\rm max}$ [km] 			& 3.5 			& $\sim 1000$	& 0.45 			& 40			\\
        \hline
        Minimum baseline, $b_{\rm min}$ [m] 		& 7.7 			& 35			& 14.6 			& 35.0		    \\
%       Maximum core baseline, $b_{\rm c, min}$ [km] & \dk 			& 3.5			& 0.292			& \dk 			\\
        Maximum baseline, $b_{\rm max}$ [km] 		& 5.3 			& 1500			& 0.879 		& 65 			\\
		Minimum baseline from \21cms, $b_{\rm min,21}$ [m]			& 7.72 			& 35.71			& 14			& 35.1			\\
		Maximum baseline from \21cms, $b_{\rm max,21}$ [km]			& 0.741			& 3.55			& 0.28			& 0.887			\\

        \hline 
        Element size [m]							& 4 			& 30.75 		& 14 			& 38 			\\
        Effective collecting area $A_e$ at 150~MHz $[{\rm m}^2]$ & 21.5 			& 512.0			& 154			& 600 			\\
		FWHM at 150~MHz [deg]						& 26			& 3.80			& 9 			& 3 			\\
		Field of view at 150~MHz [$\uni{deg}{2}$]	& 610 			& 11.35			& 64			& 12.5			\\
		Angular resolution at 150~MHz				& $2'$ 			& $3'$			& $11'$			& $5'$      	\\
		Angular resolution of core ($1.22\; \lambda/b_{\rm max,21}$)	& $11.32'$ & $2.36'$	& $29.96'$	& $9.46'$			\\
%		Angular Resolution of core from \21cms		& & & & \\
        $k_{\perp, \rm min}$ at $z = 5.5$ [$h$/cMpc]	& 0.006			& 0.027			& 0.011			& 0.027			\\
        $k_{\perp, \rm max}$ at $z = 5.5$ [$h$/cMpc]	& 0.566			& 2.711			& 0.214			& 0.677			\\

        \hline
        Minimum frequency, $\nu_{\rm min}$ [MHz] 	& 70 [167] [139]& 120 			& 50 [100] 		& 50 			\\
        Maximum frequency, $\nu_{\rm max}$ [MHz] 	& 300 [197] [167]& 240 			& 250 [200]		& 350 			\\ 
        Maximum redshift, $z_{\rm max}$ 			&19.29 [7.5] [9.2]& 10.83 		& 27.4 [13.2]	& 27.4 			\\
        Minimum redshift, $z_{\rm min}$ 			& 3.73 [6.2] [7.5]& 4.92  		& 4.7 [6.1]		& 3.06 			\\
%        Bandwidth [MHz]							& 30.72			& 32 (60)		& \dk			& \dk			\\
        Frequency (spectral) resolution [kHz]		& 40			& 61  			& 97.8			& 70    		\\
        Number of channels (in 8~MHz BW)			& 200			& 131			& 82			& 114			\\
%		LoS Comoving Resolution	(at $z = 8.5$) [Mpc]& \dk			& \dk 			& 1.7   		& \dk			\\
%		Time Resolution	[s]							& 0.5			& 2				& 10.7			& 				\\
%       Receiver Temperature [K] 					& \dk 			& \dk 			& \dk 			& \dk 			\\        
        $k_{\parallel, \rm min}$ at $z = 5.5$ [$h$/cMpc]	& 0.08      & 0.08          & 0.08          & 0.08      \\
        $k_{\parallel, \rm max}$ at $z = 5.5$ [$h$/cMpc]	& 8.2       & 5.38          & 3.35          & 4.69      \\
        
        \hline
        Latitude  									& $ 26^\circ 42'12''$S & $52^\circ 54'32''$N & $30^\circ 43'17''$S & $26^\circ 49'29''$S \\
        Longitude 									& $116^\circ 40'16''$E & $ 6^\circ 52'08''$E & $21^\circ 25'42''$E & $116^\circ 45'52''$E \\
%         & & & & \\
	\bottomrule
    \end{tabular}
    
    \begin{tablenotes}[leftmargin=*,labelindent=16pt]
      \small
        \item[a] \mwa: \cite{MWA2013,2018PASA...35...33W}; \url{https://www.mwatelescope.org/telescope}; antenna coordinates from \url{https://www.mwatelescope.org/telescope/configurations/phase-ii}
	\item[b] \lofar: \cite{LOFAR2013}; \url{http://www.lofar.org/about-lofar/system/lofar-numbers/lofar-numbers.html}
	\item[c] \hera: \cite{2016ApJ...826..181D,HERA2017}
	\item[d] \skalow: \cite{acedo2020ska}; antenna coordinates from \url{https://astronomers.skatelescope.org/wp-content/uploads/2016/09/SKA-TEL-SKO-0000422_02_SKA1_LowConfigurationCoordinates-1.pdf}
    \end{tablenotes}
    \end{threeparttable}
\end{table*}

Many ongoing and upcoming radio interferometric experiments are trying to detect the power spectrum of the 21-cm signal. \gmrt\ (Giant Metrewave Radio Telescope; \mbox{\citealt{GMRT2013}}), \lofar\ (Low Frequency Array; \mbox{\citealt{2013A&A...550A.136Y, LOFAR2013}}), \mwa\ (Murchison Widefield Array; \citealt{MWA2013, 2013PASA...30...31B, 2018PASA...35...33W}), and \paper\ (Donald C. Backer Precision Array for Probing the Epoch of Reionization; \citealt{PAPER2010}) have published upper limits for power spectrum estimates (see Section~\ref{sec:UL_1}). \hera\ (Hydrogen Epoch of Reionization Array; \citealt{HERA2017}) has started taking experimental observations while it is still under construction. \skalow\ (the low-frequency component of the Square Kilometre Array; \citealt{SKA2015b}) is planned to be operational in the next decade.

We use the publicly available code \21cms\footnote{\21cms: \url{https://github.com/jpober/21cmSense}} \citep{2013AJ....145...65P,2014ApJ...782...66P} to study the possibility of detecting the 21-cm signal with \mwa, \lofar, \hera\ and \skalow\ at various redshifts. Table~\ref{tab:Params} summarises the relevant observational parameters of these instruments. We next discuss these parameters and our choice of antenna configuration for each instrument.

%% Note that, while we calculate the sensitivity at $5\leq z \leq 6$, which corresponds to observational frequencies of 203 to 237 MHz, we have quoted the effective collecting area ($A_e$), full-width half-maximum (FWHM), field of view (FoV) and angular resolution of all antenna elements at 150 MHz, following the convention in literature.

\subsection{Interferometric Experiments}

The antenna configuration for each instrument consists of a cluster of short-baseline antennas (`core'), surrounded by a few `outrigger' or `remote' antennas. The long baselines corresponding to remote/outrigger antennas are useful for calibration and foreground removal purposes; however, for sensitivity calculations of EoR fields, only the short baselines are useful. Therefore, in our work we have considered only the core configuration for each antenna array and have ignored the remote/outrigger antennas. In Table~\ref{tab:Params}, we have listed the total number of antennas $N_a$ as number of antennas in core $N_c$ +  remote/outrigger antennas $N_r$.

% --------------- MWA element, configuration and baselines -----------
\mwa\ Phase II uses a compact configuration for EoR studies (Phase IIA). The radius for this configuration ($r_{\rm core}$) is about 300~m and it consists of 128 tiles. Of these, 72 tiles are in two hexagonal cores and 56 tiles are pseudo-randomly distributed \citep{MWA2019b}. The shortest and longest baselines for this configuration are 7.7~m and 741~m respectively, whereas the longest baseline for the complete configuration (compact + extended) is 5.3~km \citep{2018PASA...35...33W}. The compact and extended baselines correspond to a resolution (at 150~MHz) of about $11'$ and $2'$ respectively. \mwa\ tiles have an approximate side of length 4~m, corresponding to 26$^\circ$\ of FWHM at 150~MHz \citep{2016ApJ...825..114J} and a large field of view of about 610 $\uni{deg}{2}$ \citep{MWA2013}.
 
% --------------- LOFAR element, configuration and baselines -----------
\lofar\ consists of two antenna arrays: the Low Band Antennas (LBA) observe at 10--80~MHz and the High Band Antennas (HBA) observe at 120--240~MHz. We only consider the HBA array, because it covers the redshift range $z=$4.92--10.83. The 24 HBA core stations are located within a radius of around 2~km \citep{LOFAR2013}. Further, around 22 remote stations are located within the Netherlands (with baseline up to 100~km) and many international stations (with baseline up to 1500~km) are spread within the Europe. The core HBA stations are used in split mode, therefore their total number is 48, and the shortest and longest baselines are 35~m and 3.5~km respectively. \cite{LOFAR2017} have discussed that even though the shortest baseline for \lofar\ HBA is $\sim35$ m, \lofar\ HBA EoR studies have discarded the short baselines ($< 127$~m) that correspond to pairs of antennas sharing common electronics \citep{LOFAR2017,LOFAR2020}. Therefore, 100~m is often quoted in the literature as the shortest \lofar\ baseline. Long baselines ($> 250\; \lambda_o$, where $\lambda_o$ is the central wavelength of observation) are also discarded for EoR studies. In our work, we have used all baselines corresponding to all 48 core HBA stations. These are the longest baselines amongst all four instruments, giving resolution of 3' at 150~MHz.
\lofar\ HBA core, remote and international stations have different diameters, 30.75~m, 41.05~m and 56.50~m respectively; we only use one element diameter of 30.75~m in our sensitivity calculations, which corresponds to a field of view of 11.35~${\rm deg}^2$ at 150~MHz \citep{LOFAR2013}.

% --------------- HERA element, configuration and baselines -----------
The `split-core configuration' of \hera\ is planned to have 350 antenna elements, of which 320 elements will be in a densely packed hexagonal core and 30 elements will be outriggers \citep{HERA2017}. The core will be split in 3 sections (hence the name `split-core'), offset by non-integer fractions of a hex spacing. A core hexagon of 19 elements (with each side having 3 elements), had started taking observations in 2017 \citep{HERA2017}. Two outrigger antennas were added in 2019 to improve foreground imaging \citep{2019AAS...23334922M}. However, since the final antenna coordinates are not publicly available yet, in our calculations we assume a perfect hexagon with 331 elements (with each side having 11 elements) \citep{2016ApJ...826..181D} in our calculations and we have ignored the outrigger antennas. This difference of 11 antennas will not have a significant effect on our sensitivity predictions. The shortest and longest baselines for this perfect hexagon are 14~m and 280~m respectively, whereas including the outrigger antennas, the longest baselines is 879~m. This corresponds to angular resolution (at 150~MHz) of $30'$ (core) and $11'$ (outriggers). Each \hera\ element consists of a 14~m dish, with a collecting area of 155~${\rm m}^2$ and a field of view of 64~$\uni{deg}{2}$.

% --------------- SKA element, configuration and baselines -----------
With 38~m diameter and a collecting area of 600~$\uni{m}{2}$, \skalow\ will have the largest element size \citep{acedo2020ska}, with a field of view of about $12.5~\uni{deg}{2}$ at 150~MHz. The full \skalow\ configuration will extend up to a radius of 40~km and the longest baseline will be 65~km ($7.7''$ resolution at 150~MHz).
In our calculations we have used 224 core elements spread within a radius of 500~m. The shortest and longest baseline of the core is 35.1~m\footnote{Note that this length is \textit{smaller} than the planned tile diameter of 38~m \citep{acedo2020ska}.} and 887~m, respectively. This results in an angular resolution of $10'$ at 150~MHz.

% --------------- frequency -----------
% MWA
\mwa\ is designed to observe in the frequency range of 70--300~MHz, which corresponds to the redshift range of $3.73 \lesssim z \lesssim 19.29$ for $\lambda_0 = 0.21$~m. However, \mwa\ EoR observations are taken in three bands: ultralow band at 75--100~MHz ($ 13.2 \lesssim z \lesssim 18$), low band at 139--167~MHz ($ 7.5 \lesssim z \lesssim 9.2$), and high band at 167--197~MHz ($ 6.2 \lesssim z \lesssim 7.5$) \citep{2016ApJ...825..114J}.
% More than ninety per cent of data is observed in low and high bands. 
The spectral (frequency) resolution of \mwa\ is 40~kHz \citep{MWA2019b,MWA2020}. This corresponds to 200 channels in a bandwidth of 8~MHz.
% LOFAR
\lofar\ HBA is optimized for the frequency range 120--240~MHz ($4.9 \lesssim z \lesssim 10.8$); however, the EoR observations are usually carried out in three narrower bands (7.9--8.6, 8.6--9.6 and 9.6--10.6) and the frequency resolution is 61~kHz for EoR studies \citep{LOFAR2017,LOFAR2020}. 
While the \lofar\ observations can have a typical duration of 12--16~hr per day \citep{LOFAR2020}, we have only assumed 6~hr of observation duration per day for all instruments.
% HERA
For \hera\ the extended frequency range is 50--250~MHz ($4.7 < z < 27.4$) and its EoR frequency band is 100--200~MHz ($6.1 < z < 13.2$), with a frequency resolution of 97.8~kHz \citep{HERA2017}.
% SKA
\skalow\ is planned to work in the frequency range of 50--350~MHz ($3 < z < 27.4$). If this whole frequency band is available for 21-cm observations, then \skalow\ will provide us information from the formation of first stars and galaxies to the end of hydrogen reionization.
In next sub-section we present results for all four instruments in the redshift range $5<z<6$, assuming that their observational range extends to these redshifts.

\subsection{Sensitivity of 21-cm Observations}

\begin{figure*}
    \includegraphics[width=\textwidth]{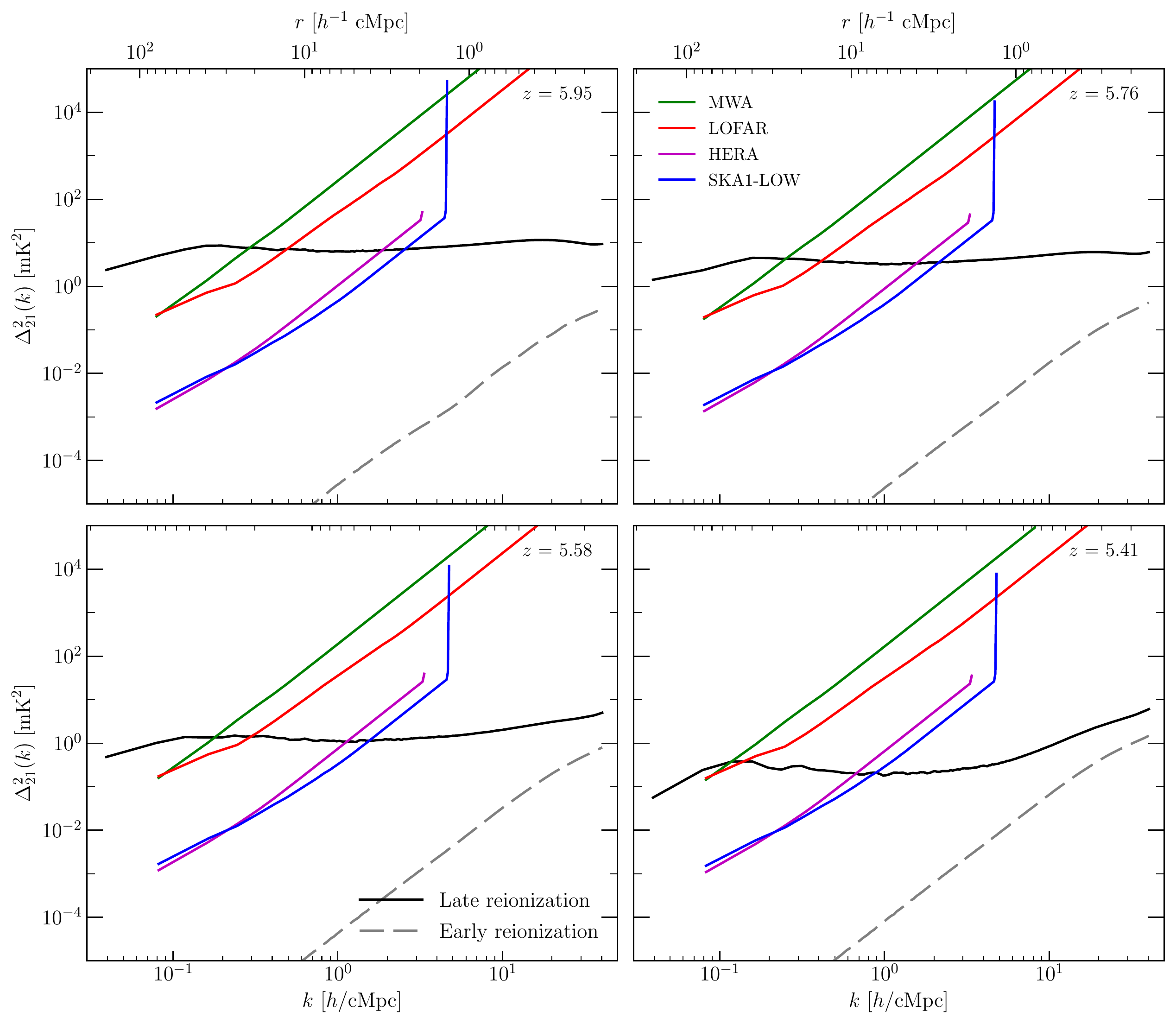}\\
    \caption{The 21-cm power spectra from our models compared to the noise power spectra for \mwa\ (green curves), \lofar\ (red curves), \hera\ (magenta curves), and \skalow\ (blue curves), for tracking mode observations of 6 hours a day, 180 days (total 1080 hours) at $z=5.94$ ($\nu = 204$~MHz; top left panel), $z=5.76$ ($\nu = 210$~MHz; top right panel), $z=5.58$ ($\nu = 216$~MHz; bottom left panel) and $z=5.41$ ($\nu = 221$~MHz; bottom right panel). Black solid curves correspond to power spectra from our preferred late reionization model.  Grey dashed curves show predictions for an early reionization  model.}
  \label{fig:sense_all}
\end{figure*}

In this subsection, we discuss the \textsc{21cmSense} code used to calculate the sensitivity of various observational instruments and present the comparison for 1080~hr of observation (6~hr of tracking mode observation per day, for 180~days). 
Sensitivity of an instrument indicates how weak a signal can be detected by that instrument.  
In the absence of the signal, the noise detected by the instrument is a Gaussian random variable with zero mean.
Hence, statistically independent observations can be combined to improve upon the sensitivity $\Delta_{N,0}^2(k)$. This improvement is inversely proportional to the square root of the number of samples $N_s$ ($\Delta_N^2(k) = \Delta_{N,0}^2(k)/\sqrt{N_s}$).

Given an antenna configuration for an interferometer and the wavelength of observation $\lambda (z)$, the baseline distribution is calculated by \textsc{21cmSense} using $u_{ij} (z) = b_{ij}/\lambda(z)$, where $b_{ij}$ is the physical distance between a pair of antennas $i$ and $j$. %The auto-correlation of antenna pairs is ignored ($i \neq j$). 
While it is possible to specify the minimum and maximum baseline length while using \21cms for sensitivity calculations, we use all the baselines provided by the antenna configuration in our calculations (see discussion above). Given $N_a$ number of antennas, the total number of baselines is $N_a(N_a-1)/2$. As each baseline is an independent measurement, the number of data samples increases with the number of baselines ($N_s \propto N_a^2$).

% $\mathd u = \mathd ({\rm sin}(\theta)) \simeq 1/\mathd \theta$
The angle extended over the sky by a baseline $u$ is $\theta \simeq 1/u$. This approximation is only valid when the angle $\theta$ is small (small angle approximation). The transverse distance in the sky extended by angle $\theta$ at redshift $z$ is \citep{Cosmo_low},
\begin{align}
	l_\perp (z) = X(z)  \theta & \approx 1.9\,{\rm cMpc}/h \left(\frac{1+z}{10}\right)^{0.2}\! \left(\frac{\theta}{1\,\mathrm{arcmin}}\right).
\end{align}
Therefore, $k$ mode of the power spectrum perpendicular to the line-of-sight ($k_\perp$) at redshift $z$ for  $u \gg 1$  is $k_\perp (z) \simeq 2\pi/l_\perp (z) \simeq (2\pi/X(z)) u$. The smallest and largest $k_\perp$ modes probed at any redshift are determined by the smallest and largest baseline lengths respectively ($b_{\rm min}$ and $b_{\rm max}$).
At $z= 5.5$, $X(z) = 1.743~{\rm cMpc}/h/{\rm arcmin}$ and $\lambda_o = 1.37~{\rm m}$. Therefore for a baseline of 10~m, $u_{10{\rm m}}(5.5) = 7.28$ and the transverse $k$-mode probed is, $k_{\perp, 10{\rm m}} (5.5) \sim 0.0076~{\rm cMpc}/h$. For a 100~m baseline, $k_{\perp, 100{\rm m}} (5.5) \sim 0.076~{\rm cMpc}/h$.  

%$X$ and $Y$ are cosmological constants.
Ignoring redshift-space distortions, the line of sight distance at redshift $z$ covered by observational frequency range $\Delta \nu$ is \citep{Cosmo_low}, 
\begin{align}
	 & l_\parallel (z) = Y(z) \Delta \nu \nn \\
	 & ~\approx 11.5\,\mathrm{cMpc}/h \left(\frac{1+z}{10}\right)^{0.5}\! \left(\frac{\Omega_m h^2}{0.15}\right)^{-0.5}\!\! \left(\frac{\Delta \nu }{1\,\mathrm{MHz}}\right),
\end{align}
and the line of sight $k$-mode at redshift $z$ is $k_\parallel (z) = (2\pi/Y(z)) \eta$, where the delay parameter $\eta$ is the Fourier transform of the frequency range $\Delta \nu$.
The smallest and largest $k_\parallel$ modes probed are determined by the cosmological bandwidth $B$ and channel width $\Delta \nu_c$ respectively. Here $B$ is the redshift range that can be considered cosmologically co-eval. We have used 8~MHz as default value for the bandwidth in our calculations using \textsc{21cmSense}. The number of samples increases with the bandwidth as $N_s \propto B$. Therefore, $\Delta_N^2(k) \propto \Delta_{N,0}^2(k)/\sqrt{B}$. 
The number of channels in the bandwidth ($n_{\rm chan} = B/\Delta\nu_c$) is a function of the instrument (see Table~\ref{tab:Params}). It affects the maximum $k$-mode that can be probed, but does not have a significant effect on the sensitivity.
% Y(5.5) = 9.56 cMpc/h/MHz

If the total number of days observed is $n_{\rm days}$ 
and the number of observing hours per day is $t_\text{per-day}$, then the sensitivity of the instrument is \citep{2012ApJ...753...81P},
\begin{align}
	\Delta^2_{N}(k) &\approx  \frac{X^2Y}{4\pi^2} [k]^{\frac{5}{2}} [\Omega] \left[\frac{1}{t_\text{per-day}}\right]^{\frac{1}{2}} \left[\frac{1}{n_{\rm days}}\right] \left[\frac{1}{B}\right]^{\frac{1}{2}} \nonumber \\
	& \qquad \left[\frac{1}{\Delta \;{\rm ln}(k)}\right]^{\frac{1}{2}}\left[\frac{1}{N_a}\right] \left[\frac{f_0}{f}\right]^{\frac{1}{2}} \; T_{\rm sys}^2 (u,v,\eta).
	\label{eq:sensitivity}
\end{align}
Here, $\Omega$ is the primary beam field of view. It is assumed to be a 2D Gaussian for all instruments. $f$ and $f_0$ are baseline redundancy parameters, which we discuss below. $\Delta \;{\rm ln} (k)$ is the number of $k$-modes in a logarithmic bin.

We have used  the `moderate' model for the foreground wedge, where all $k$-modes inside the horizon are excluded from the sensitivity calculations and all baselines within a $uv$-pixel are added coherently. Purely north-south baselines are excluded from the calculations. 
The system temperature has contributions from both the sky and the instrument ($T_{\rm sys} = T_{\rm sky} + T_{\rm rec}$). We have taken the receiver temperature to be $T_{\rm rec} = 100\; {\rm K}$ for all instruments and the sky temperature is taken to be \citep{2001isra.book.....T},
\begin{align}
    T_{\rm sky} = 60\;{\rm K} \left(\frac{300\;{\rm MHz}}{\nu}\right)^{2.55}. \label{eq:Tsky}
\end{align}
The sky temperature decreases with increasing frequency $\nu$, because it is set by synchrotron foregrounds, which are weaker at higher frequencies.
$T_{\rm sys}$ is orders of magnitude larger than the 21-cm signal (Equation~\ref{eq:21Tb}), therefore the first-generation 21-cm observations are expected to have poor signal to noise ratio ($S/N<1$). Instruments planning to observe the 21-cm signal have been designed to redundantly sample baselines \citep{2016ApJ...826..181D}. 
If a single baseline with a $t_0$ integration time and $n_i$ number of $t_0$ samples in $uv$-bin $i$ has $f_0$ sampling redundancy, then the increased sensitivity for a redundant array is $ f \equiv f_0 \left(\sum_i n_i^2/\sum_i n_i \right) $ \citep{2012ApJ...753...81P}.
% Therefore, for a redundant antenna array configuration, $\Delta_N^2(k) \propto (1/N_a) (f_0/f)^{1/2}  \Delta_{N,0}^2(k)$.

% NOTE: \cite{HERA2017} gives system temperature as, $T_{\rm sys}= 100 + 120\;(\nu/150\; {\rm MHz})^{-2.55}\; {\rm K}$, which (almost) matches with the our Eq~\ref{eq:Tsky}.

% NOTE: 180 is the maximum number of days a particular R.A. can be observed in one year observations are only taken at night {source - 21cmsense}.

\21cmsense uses a detailed version of Equation~\ref{eq:sensitivity} to calculate instrument sensitivity for a given configuration.
In Figure~\ref{fig:sense_all}, we show our sensitivity predictions for \mwa, \lofar, \hera\ and \skalow\ for redshifts $5 < z < 6$ for 180~days and 6~hours of tracking mode observation duration per day. The total observing time is $n_{\rm days} \times t_\text{per-day} = 1080$~hours. We have shown our late reionization power spectrum prediction in this range, along with the early reionization power spectra. For our late reionization model, \hera\ and \skalow\ will be able to detect the signal at $5.4 \leq z \leq 6$, in 1080~hr of observation and clear distinction between late and early reionization models can be made in this redshift range. \mwa\ and \lofar\ HBA can only detect signal at low $k$-modes at $z>5.4$. 

The minimum value of the $k_\parallel$ and $k_\perp$ modes depend on the bandwidth (8~MHz) and the minimum baseline of the instrument, respectively. 
The maximum value of $k_\parallel$ and $k_\perp$ modes are determined by channel width and maximum baseline (maximum `core' baseline in our case) of the instrument, respectively. Here, the smaller frequency resolution of \mwa\ and the large baselines of \lofar\ lead to extended sensitivity at large $k$-modes. In comparison, the large frequency resolution and short core baselines of \hera\ result in much lower cutoff of $k$ modes. 
As $k_{\perp, {\rm max}}$ increases linearly with the maximum baseline length, assuming the full configuration for \mwa\ and \skalow\ instead of the `core' antenna configuration would increase $k_{\perp, {\rm max}}$ by an order of magnitude (Table~\ref{tab:Params}). For \lofar, taking into account the international baselines can theoretically lead to probing $k_\perp$ of $10^3~h$/cMpc. 
However, large baselines are not very useful for detecting the 21cm signal as the sensitivity decreases rapidly for high $k$-modes (Equation~(\ref{eq:sensitivity}) and Figure~\ref{fig:sense_all}). Hence, none of the instruments will be able to probe the signal at these scales at redshifts of interest.

Since most of the reionization models used in the past predict an end of reionization by $ z \sim 6$, these inteferometric instruments do not plan to observe at redshifts below 6. For example, the high band of \mwa\ only observes $6.2 \leq z \leq 7.5$ and the EoR band of \hera\ will observe $6.1 \leq z \leq 13.2$. However, if the reionization of \hi ends somewhat later, then there is two to four orders of magnitude more power in 21-cm spectra at $z<6$ than in, e.g., the early reionization model, and extending the observation frequency of \hera\ and \skalow\ to these redshifts will help to make a clear distinction between these models. It is also important to note that while some instruments (e.g., \citealt{MWA2014}) plan to observe post-reionization frequencies to measure the `null signal' due to residual foregrounds, these redshifts might still contain strong 21-cm signal, if reionization ends later than $z\sim 6$. 

% NOTE: Dillon et al. 2014 -- "a broad frequency range—with access to a wide range of redshifts, future detections of the cosmological signal can be distinguished from residual foregrounds by measuring null signals at redshifts where reionization is complete."

\subsection{Current Upper Limits}\label{sec:UL_1}
\begin{figure*}
  \includegraphics[width=\textwidth]{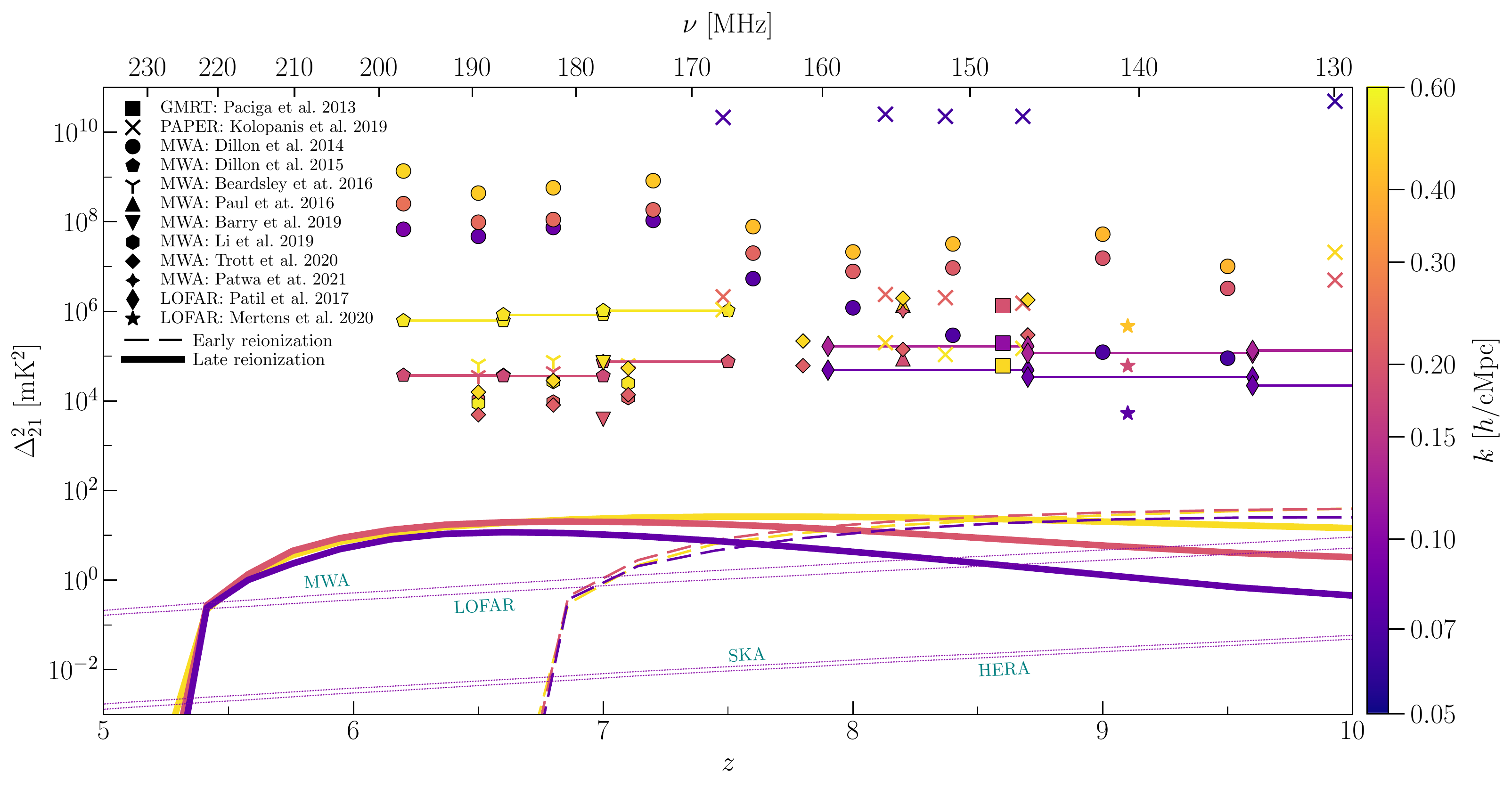}
  \caption{Upper limits on the 21-cm power spectrum $\Delta_{21}^2$ reported by various interferometeric experiments compared with the theoretical predictions of the power spectrum values of our simulation at $k=0.079~h$/cMpc (yellow solid curve), $k=0.197~h$/cMpc (orange solid curve) and $k=0.512~h$/cMpc (purple solid curve). The dashed curves show the power spectrum evolution in an early reionization model as discussed in the text.  When upper limits are not available at these $k$ values, limits at the closest available $k$ value are shown.  The numerical data compilation is presented in Appendix~\ref{sec:UL}. Instrument sensitivities of \mwa, \lofar, \hera\ and \skalow\ for 1080~hr of observations at $k=0.1~h$/cMpc are shown as dotted purple lines.}
  \label{fig:UL}
\end{figure*}

In Figure~\ref{fig:UL} we show the power spectrum from the late and early reionization models at $k=0.079$, 0.197 and 0.512~$h$/cMpc. The colour scheme represents a range of $k$-modes from 0.05 to 0.6~$h/${\rm cMpc}.  Our simulated late reionization model has significant power up to redshift $z \sim 5.4$. At $z < 6.4$, these power spectra vary mildly with $k$ (see also Figures~\ref{fig:PS_all2} and~\ref{fig:sense_all}).  At small scales (large $k$), our power spectra have large power at all redshifts due to the small-scale structures in the IGM.  However, at large scales (small $k$) the power increases with time as the size of ionized regions increases. Power at all scales declines rapidly towards the end of reionization ($z \lesssim 6$); this corresponds to $x_{\rm HII} \gtrsim 0.8$ in Figure~\ref{fig:ave_all}. For $k=0.2~h$/cMpc, the peak of the power spectra seems to coincide with the midpoint of reionization at $z \sim 7.1$, whereas for larger scales, $k=0.08~h$/cMpc, the peak of the power spectra is delayed to $z\sim 6.5$ when the ionized regions grow due to percolation in this model. For the early reionization model, the power at all scales is comparable and the evolution due to redshift is negligible. Note that we have ignored the fluctuations due to inhomogeneous spin temperature in this paper ($T_{\rm S} \gg T_{\rm CMB}$). While the fluctuations due to inhomogeneous \lyal coupling and heating are expected to be small at $z<10$ \citep{Ghara2015,2017MNRAS.472.1915C}, delayed heating due to inefficient X-ray sources or low escape fraction of X-ray photons can enhance the power spectra by up to two orders of magnitude \citep{LateHeat2, 2014MNRAS.439.3262M,RS18, RS19,2021ApJ...912..143M}. Therefore, other than the redshift of heating transition ($T_{\rm K} \simeq T_{\rm CMB}$, when power at large scale decreases), our power spectra represent conservative limits for their respective reionization model.
We hope to analyse the effect of inhomogenous $T_{\rm S}$ on our late reionization model predictions in future work.

While the Ly$\alpha$-informed models predict strong 21-cm power spectrum signal at relatively lower redshifts, these low redshifts are also potentially more convenient for experiment.  We show our sensitivity predictions for \mwa, \lofar, \hera\ and \skalow\ as function of redshift for 1080~hr of observation at $k=0.1~h$/cMpc as thin dotted purple curves in Figure~\ref{fig:UL}. We see that \hera\ and \skalow\ have sensitivity comparable at this scale and they will be able to detect the 21-cm signal at $5.4<z<10$ in about $<1000$~hr of observation. \lofar\ and \mwa\ have sensitivities that are worse by about two orders of magnitude. They will be able to detect the 21-cm signal at redshifts around the midpoint of reionization where the 21-cm signal has large power due to fluctuations of the ionization field. All four instrument sensitivities increase at higher redshifts due to increasing sky temperature (Equation~\ref{eq:Tsky}). For example, the sensitivities increase by an order of magnitude from redshift 5.5 to redshift 8.

So far most 21-cm epoch-of-reionization experiments have focused exclusively at $z>6$.  This is evident in the upper limits on the 21-cm power spectrum reported by these experiments.  In Figure~\ref{fig:UL}, we compile several upper limits on the 21-cm power spectrum $\Delta_{21}^2$.  This includes limits published  by \gmrt\ \citep{GMRT2013}, \lofar\ \citep{LOFAR2017,LOFAR2020}, \mwa\ \citep{MWA2014,MWA2015,MWA2016,MWA2016b,MWA2019,MWA2019b,MWA2020, MWA2021} and \paper\ \citep{PAPER2019} at redshifts $z \leq 10$ at $k$-modes closest to 0.08, 0.20 and 0.50~$h$/cMpc. Appendix~\ref{sec:UL} lists the numerical values of the points in Figure~\ref{fig:UL}.  When upper limits are not available at these $k$ values, limits at the closest available $k$ value are shown.  All published limits are at $z>6$.  While \lofar\ HBA works in the frequency range 120--240~MHz, which corresponds to $4.9 <z < 10.8$, their upper limits for $\Delta_{21}^2$ are given at $7.9 < z< 10.6$ for a wide range of $k$-modes (0.053~$h$/cMpc $<k<$ 0.432~$h$/cMpc) \citep{LOFAR2017,LOFAR2020}. These upper limits from \lofar\ at $7.9 < z< 10.6$ are better at smaller $k$ values (large scales). However, at these redshifts the simulated power spectra also decrease rapidly with $k$. As a result, the upper limits are still about four orders of magnitude higher than the simulation predictions at these redshifts. At $z \sim 8.6$, all four experiments have published upper limits \citep{GMRT2013,LOFAR2017,PAPER2019,MWA2020}, which are currently roughly at the same level. The upper limit provided by \cite{GMRT2013} at $z \sim 8.6$ is less than three orders of magnitude larger than our predicted signal. Upper limits from MWA at  $z \sim 8.6$ are also of the similar values \citep{MWA2016b, MWA2020, MWA2021}.
The minimum redshift studied is $z = 6.5$ by \mwa\ at 0.14~$h$/cMpc $<k<$ 0.6~$h$/cMpc \citep{MWA2015,MWA2016,MWA2019b,MWA2020}. 
They have given the current best limits of $\Delta_{21}^2$ at $ 6.5 <z < 7.8$. 
These limits are about two orders of magnitude higher than the predicted signal by our late reionization model. It is interesting to note that the difference between the current upper limits and our model prediction decreases rapidly towards lower redshifts.

\section{Conclusions} \label{sec:conclusion}

Most 21-cm experiments work under the assumption that the epoch of reionization ends at $z \sim 6$.  However, radiative transfer simulations that agree with the Ly~$\alpha$ data prefer delayed reionization.  These models suggest that neutral hydrogen islands of sizes of up to 100 comoving Mpc may persist in the IGM at redshift as low as $z \sim 5.3$. These islands can explain the spatial fluctuations seen in the \lyal forest opacity at $z<6$ \citep{2019MNRAS.485L..24K, 2020MNRAS.491.1736K}. We show in this paper that due to the presence of these large patches of neutral hydrogen, the power spectrum of 21-cm brightness temperature is significantly enhanced at redshifts $5<z<6$ relative to previous models. We compare the 21-cm power spectra at $z<6$ with a more conventional reionization model, in which reionization ends at $z > 6$, and find that there is about two to four orders of magnitude difference in the 21-cm power spectra from these two models at these redshifts.

The larger power spectra predicted by our delayed reionization model should be observable at high significance by \hera\ and \skalow\ with observation duration of 180 days with 6 hours per day (total 1080 hours), assuming optimistic foreground subtraction. To achieve a similar sensitivity, \mwa\ and \lofar\ will need to observe for about hundred times longer.  A prediction of an enhanced 21-cm power spectrum is good news for interferometric experiments as at low redshifts (high frequencies) the thermal noise due to foregrounds is considerably lower ($T_{\rm sky} \propto \nu_c^{-2.55}$) and the sensitivity of the instruments is correspondingly better.  This is worth noting for experimental efforts that have been artificially restricted to $z>6$. In particular, it might be worthwhile for \hera\ and \skalow\ to plan epoch-of-reionization observations at $z<6$. 

Another benefit of relatively lower redshifts is better synergies with
multi-wavelength experiments.  Over the next few years, optical/IR
facilities such as the James Webb Space Telescope (\textit{JWST}), the
Vera C. Rubin Observatory, the Nancy Grace Roman Space Telescope, and
Euclid space telescope, in addition to \textsc{alma} and the
thirty-metre-class telescopes, will provide data at $z\sim 6$ that
will potentially identify sources of reionization.  The Ly~$\alpha$
emitter clustering and luminosity function measurements from
facilities such as Subaru/HSC and Subaru/PFS are also available at
$z<7.5$ \citep{2020MNRAS.494..703W, 2019MNRAS.485.1350W}.  Metal-line
intensity mapping experiments such as \textsc{concerto}
\citep{2020A&A...642A..60C} and \textsc{ccat-p}
\citep{2020JLTP..199..898C} will also potentially detect the
large-scale structure at these redshifts.  Cross-correlating 21-cm
measurements with these multi-wavelength data sets can potentially
yield important scientific insight by reducing parameter degeneracies
\citep{2019MNRAS.485.3486D}.

\section*{Acknowledgements}

We thank Somnath Bharadwaj, Tirthankar Roy Choudhury, Avinash
Deshpande, L\'eon Koopmans, Akash Kumar Patwa, and Saurabh Singh for
useful comments.  This work was supported by a grant from the Swiss
National Supercomputing Centre (CSCS) under project ID s949.  GK
gratefully acknowledges support by the Max Planck Society via a
partner group grant.  This work used the Cambridge Service for Data
Driven Discovery (CSD3) operated by the University of Cambridge
(www.csd3.cam.ac.uk), provided by Dell EMC and Intel using Tier-2
funding from the Engineering and Physical Sciences Research Council
(capital grant EP/P020259/1), and DiRAC funding from the Science and
Technology Facilities Council (www.dirac.ac.uk).  This work further
used the COSMA Data Centric system operated Durham University on
behalf of the STFC DiRAC HPC Facility. This equipment was funded by a
BIS National E-infrastructure capital grant ST/K00042X/1, DiRAC
Operations grant ST/K003267/1 and Durham University. DiRAC is part of
the UK's National E-Infrastructure.

\section*{Data Availability}

No new data were generated or analysed in support of this research.
                                                                                                       
%%%%%%%%%%%%%%%%%%%%%%%%%%%%%%%%%%%%%%%%%%%%%%%%%%


\bibliographystyle{mnras}

%%%%%%%%%%%%%%%%% APPENDICES %%%%%%%%%%%%%%%%%%%%%

\appendix

\section{21-cm power spectrum upper limits}\label{sec:UL}
In Figure~\ref{fig:UL}, we have compared the 21-cm brightness temperature power spectrum from our model with a model of early reionization at $k=0.08\,h$/cMpc, $0.2\,h$/cMpc and $0.5\,h$/cMpc. We also compare these model predictions with a compilation of 21-cm power spectrum upper limits reported by \textsc{gmrt} \citep{GMRT2013}, \lofar\ \citep{LOFAR2017,LOFAR2020}, \mwa\ \citep{MWA2014,MWA2015,MWA2016,MWA2016b,MWA2019,MWA2019b,MWA2020,MWA2021} and \paper\ \citep{PAPER2019}. The numerical values of these limits are listed in Table~\ref{tab:UL}.

\begin{table}
	\centering
	\caption{A compilation of the 21-cm power spectrum upper limits reported by \textsc{gmrt}, \lofar, \mwa\ and \paper\ at redshifts $5 \leq z \leq 10$ at $k$-modes close to 0.08~$h$/cMpc, 0.2~$h$/cMpc and 0.5~$h$/cMpc. Figure~\ref{fig:UL} compares these values with our predictions.}
	\label{tab:UL}
    \begin{tabular}{c|c|c}
	\toprule
		$z$ & $k$ [$h/ \rm cMpc$] & $\Delta_{21}(k) \; [{\rm mK}] $ \\
		\hline
		\multicolumn{3}{c}{\textbf{\gmrt}: \cite{GMRT2013}} \\
		\hline
		8.6 & 0.10 & 443	\\
%		8.6 & 0.16 & 318	\\
		8.6 & 0.19 & 1156	\\
		8.6 & 0.50 & 248	\\
		\hline
        \multicolumn{3}{c}{\textbf{\lofar}: \cite{LOFAR2017}} \\
		\hline
		7.9--8.6	& 0.083 & 220.9	\\
		7.9--8.6	& 0.128 & 407.7	\\ 
		8.6--9.6	& 0.083 & 184.7	\\
		8.6--9.6	& 0.128 & 342.0	\\
		9.6--10.6	& 0.083 & 148.6	\\
		9.6--10.6	& 0.128 & 366.1	\\
		\hline
        \multicolumn{3}{c}{\textbf{\lofar}: \cite{LOFAR2020}} \\ %NOTE: This is in the redshift range $8.7<z<9.6$
		\hline
        9.1 & 0.075 & 72.86		\\
		9.1 & 0.179 & 246.92	\\ 
		9.1 & 0.432 & 683.20	\\    
		\hline
		\multicolumn{3}{c}{\textbf{\mwa}: \cite{MWA2014}} \\
		\hline
		6.2 & 0.093 & $ 8.25 \times 10^3 $ \\
		6.2 & 0.246 & $ 1.59 \times 10^4 $ \\
		6.2 & 0.487 & $ 3.69 \times 10^4 $ \\
		6.5 & 0.081 & $ 6.88 \times 10^3 $ \\
		6.5 & 0.233 & $ 9.89 \times 10^3 $ \\
		6.5 & 0.457 & $ 2.09 \times 10^4 $ \\
		6.8 & 0.084 & $ 8.63 \times 10^3 $ \\
		6.8 & 0.233 & $ 1.06 \times 10^4 $ \\
		6.8 & 0.451 & $ 2.40 \times 10^4 $ \\
		7.2 & 0.080 & $ 1.03 \times 10^4 $ \\
		7.2 & 0.225 & $ 1.36 \times 10^4 $ \\
		7.2 & 0.446 & $ 2.87 \times 10^4 $ \\
		7.6 & 0.074 & $ 2.31 \times 10^3 $ \\
		7.6 & 0.233 & $ 4.47 \times 10^3 $ \\
		7.6 & 0.426 & $ 8.83 \times 10^3 $ \\
		8.0 & 0.073 & $ 1.10 \times 10^3 $ \\
		8.0 & 0.214 & $ 2.79 \times 10^3 $ \\
		8.0 & 0.424 & $ 4.61 \times 10^3 $ \\
		8.4 & 0.070 & $ 5.40 \times 10^2 $ \\
		8.4 & 0.206 & $ 3.06 \times 10^3 $ \\
		8.4 & 0.419 & $ 5.66 \times 10^3 $ \\
		9.0 & 0.069 & $ 3.50 \times 10^2 $ \\
		9.0 & 0.200 & $ 3.93 \times 10^3 $ \\
		9.0 & 0.405 & $ 7.27 \times 10^3 $ \\
		9.5 & 0.067 & $ 3.00 \times 10^2 $ \\
		9.5 & 0.197 & $ 1.80 \times 10^3 $ \\
		9.5 & 0.399 & $ 3.18 \times 10^3 $ \\
		\hline
		\multicolumn{3}{c}{\textbf{\mwa}: \cite{MWA2015}} \\
		\hline
%        $3.7 \times 10^4$ & 0.18 & 6.8 \\
		6.2--6.6 & 0.181 & $ 1.93 \times 10^2$ \\ 
		6.2--6.6 & 0.537 & $ 7.87 \times 10^2$ \\ 
		6.6--7.0 & 0.176 & $ 1.90 \times 10^2$ \\
		6.6--7.0 & 0.526 & $ 9.17 \times 10^2$ \\
		7.0--7.5 & 0.195 & $ 2.74 \times 10^2$ \\
		7.0--7.5 & 0.550 & $ 1.02 \times 10^3$ \\
	\bottomrule
	\end{tabular}
\end{table}

\begin{table}
	\centering
	\contcaption{21-cm power spectrum upper limits}
    \begin{tabular}{c|c|c}
	\toprule
		$z$ & $k$ [$h/$cMpc] & $\Delta_{21}(k)\; [{\rm mK}] $ \\
		\hline
		\multicolumn{3}{c}{\textbf{\mwa}: \cite{MWA2016}} \\
		\hline
        6.5 & 0.20 & 193.80		\\
		6.5 & 0.53 & 259.29 	\\
		6.8 & 0.20 & 213.57 	\\
		6.8 & 0.52 & 284.80 	\\
		7.1 & 0.20 & 215.44 	\\
		7.1 & 0.51 & 263.30 	\\
		\hline
		\multicolumn{3}{c}{\textbf{\mwa}: \cite{MWA2016b}} \\
		\hline
		8.2 & 0.184 & 295.3		\\
		8.2 & 0.462 & 1175.6	\\
		\hline
		\multicolumn{3}{c}{\textbf{\mwa}: \cite{MWA2019}} \\
		\hline
        7.0 & 0.200 & 62.37		\\ 
		7.0 & 0.523 & 266.27	\\ 
		\hline
		\multicolumn{3}{c}{\textbf{\mwa}: \cite{MWA2019b}} \\
		\hline
        6.5 & 0.222 & 102.96	\\
		6.5 & 0.554 & 93.59		\\
		6.8 & 0.217 & 97.01		\\
		6.8 & 0.542 & 164.62	\\
		7.1 & 0.212 & 109.09	\\
		7.1 & 0.531 & 157.48	\\
		\hline
		\multicolumn{3}{c}{\textbf{\mwa}: \cite{MWA2020}} \\
		\hline
		6.5 & 0.212 & 70.20		\\ 
		6.5 & 0.495 & 125.50	\\ 
		6.8 & 0.212 & 90.00 	\\ 
		6.8 & 0.495 & 169.00	\\ 
		7.1 & 0.212 & 117.40	\\
		7.1 & 0.495 & 231.90	\\ 
		7.8 & 0.212 & 247.50	\\
		7.8 & 0.495 & 466.80	\\ 
		8.2 & 0.212 & 376.30	\\
		8.2 & 0.495 & 1402.60 	\\ 
		8.7 & 0.212 & 544.70	\\
		8.7 & 0.495 & 1341.00	\\ 
		\hline
		\multicolumn{3}{c}{\textbf{\mwa}: \cite{MWA2021}} \\
		\hline
		8.2 & 0.20 & $1000$   \\ 
		\hline
        \multicolumn{3}{c}{\textbf{\paper}: \cite{PAPER2019}} \\
    	\hline
        7.48 & 0.069 & $ 1.46 \times 10^5 $	\\
        7.48 & 0.231 & $ 1454.36 $			\\
		7.48 & 0.506 & $ 1044.32 $			\\
		8.13 & 0.066 & $ 1.59 \times 10^5 $	\\ 
		8.13 & 0.223 & $ 1543.91 $			\\ 
		8.13 & 0.488 & $ 446.38 $			\\ 
		8.37 & 0.065 & $ 1.50 \times 10^5 $	\\ 
		8.37 & 0.220 & $ 1421.33 $			\\ 
		8.37 & 0.534 & $ 329.58 $			\\ 
		8.68 & 0.064 & $ 1.50 \times 10^5 $	\\ 
		8.68 & 0.217 & $ 1232.40 $			\\ 
		8.68 & 0.525 & $ 386.66 $			\\
		9.93 & 0.060 & $ 2.21 \times 10^5 $	\\ 
		9.93 & 0.204 & $ 2231.54 $			\\ 
		9.93 & 0.495 & $ 4564.18 $			\\ 
	\bottomrule
	\end{tabular}
\end{table}

%%%%%%%%%%%%%%%%%%%%%%%%%%%%%%%%%%%%%%%%%%%%%%%%%%

% Don't change these lines
\bsp	% typesetting comment
\label{lastpage}
\end{document}